\documentclass[sigconf]{acmart}
\usepackage{cleveref}

\usepackage[most]{tcolorbox}
\usepackage{xcolor}
\usepackage{listings}
\usepackage{fontawesome}
\usepackage{ulem}

\newtcolorbox{findingbox}{
    colback=blue!10,
    colframe=blue!10,
    boxrule=0pt,
    arc=10pt,
    left=5pt,
    right=5pt,
    top=0pt,
    bottom=0pt,
    width=\linewidth,
}

\newtcolorbox{promptlistingbox}{
    colback=blue!10,
    colframe=blue!10,
    boxrule=0pt,
    arc=10pt,
    left=5pt,
    right=5pt,
    top=5pt,
    bottom=5pt,
    width=\linewidth,
    enhanced,
    before skip=10pt,
    after skip=10pt,
}

\lstdefinestyle{prompt}{
    basicstyle=\ttfamily\small,
    breaklines=true,
    backgroundcolor=\color{blue!10},
    frame=none,
    showstringspaces=false,
}

\tcbset{colback=gray!5!white, colframe=gray!80!black, boxrule=0.5pt, arc=2pt, left=6pt, right=6pt, top=1pt, bottom=1pt}
\tcbset{
  resultbox/.style={
    colback=blue!5!white,
    colframe=blue!75!black,
    title font=\bfseries,
    coltitle=white,
    boxrule=0.5mm,
    arc=2mm,
    left=2mm,
    right=2mm,
    top=1mm,
    bottom=1mm
  }
}

\newcommand{\finding}[2]{%
  \begin{findingbox}
    \textbf{Synthesis \##1.} \textit{#2}
  \end{findingbox}
}

\AtBeginDocument{%
  }

\setcopyright{licensedothergov}
\copyrightyear{2025}
\acmYear{2025}
\acmDOI{10.XXXX/0000000.0000000}

\acmConference[SBES 2025]{XXXVIII Brazilian Symposium on Software Engineering}{October,
  2025}{PE}

\setcopyright{none} 
\renewcommand\footnotetextcopyrightpermission[1]{}
\acmBooktitle{XXXVIII Brazilian Symposium on Software Engineering (SBES 2025), October, 2025, Recife, Brazil}
\acmPrice{0.00}
\acmISBN{0000-8-0000-000-2/00/12}

\begin{document}

\title{Generating Proto-Personas through Prompt Engineering: A Case Study on Efficiency, Effectiveness and Empathy}

\author{Fernando Ayach, Vitor Lameirão, Raul Leão, 
Jerfferson Felizardo, Rafael Sobrinho, Vanessa Borges, Patrícia Matsubara, Awdren Fontão}
\affiliation{
  \institution{Faculty of Computing - Federal University of Mato Grosso do Sul}
  \city{Campo Grande}
  \state{MS}
  \country{Brazil}
}
\email{{fernando.ayach, vitor.lameirao, raul.leao, jerfferson.junior, rafael.r.sobrinho }@ufms.br}
\email{{ vanessa.a.borges, patricia.gomes, awdren.fontao}@ufms.br}

\renewcommand{\shortauthors}{Ayach et al.}

\begin{abstract}
Proto-personas are commonly used during early-stage Product Discovery, such as Lean Inception, to guide product definition and stakeholder alignment. However, the manual creation of proto-personas is often time-consuming, cognitively demanding, and prone to bias. In this paper, we propose and empirically investigate a prompt engineering-based approach to generate proto-personas with the support of Generative AI (GenAI). Our goal is to evaluate the approach in terms of efficiency, effectiveness, user acceptance, and the empathy elicited by the generated personas. We conducted a case study with 19 participants embedded in a real Lean Inception, employing a qualitative and quantitative methods design. The results reveal the approach’s efficiency by reducing time and effort and improving the quality and reusability of personas in later discovery phases, such as Minimum Viable Product (MVP) scoping and feature refinement. While acceptance was generally high, especially regarding perceived usefulness and ease of use, participants noted limitations related to generalization and domain specificity. Furthermore, although cognitive empathy was strongly supported, affective and behavioral empathy varied significantly across participants. These results contribute novel empirical evidence on how GenAI can be effectively integrated into software Product Discovery practices, while also identifying key challenges to be addressed in future iterations of such hybrid design processes.
\end{abstract}




\maketitle

\section{Introduction}

Software Engineering (SE) teams are increasingly pressured to deliver productivity and quality while dealing with complexity and cost constraints \cite{banh2025copiloting}. In this context, Product Discovery plays a key role in Requirements Engineering (RE) by helping teams rapidly understand user needs and define initial development directions \cite{trieflinger2023discovery}. Proto-personas are early user representations based on assumptions and limited data, commonly used to guide product development before thorough research \cite{gothelf2013lean}. However, creating proto-personas is often a subjective and error-prone task, especially when teams have limited experience in user research or operate under strict time and resource constraints \cite{pinheiro2018contribution}. In such cases, proto-personas often rely on intuition or internal bias.

Early-stage design has seen limited integration with emerging technologies like GenAI. Context-sensitive and empathetic tasks, such as requirements elicitation and persona modeling, remain difficult to automate \cite{jackson2025exploring, sauvola2024future}. GenAI shows promise in RE tasks due to its advanced reasoning and natural language interaction capabilities \cite{marques2024using,banh2025copiloting}, particularly in supporting creative processes such as idea generation and proto-persona design. However, its adoption remains limited in ambiguous, user-driven contexts. It is influenced not only by technological and organizational factors \cite{russo2024navigating}, but also by concerns about reliability, frequent need for human oversight \cite{jackson2025exploring}, and the risk that excessive automation may compromise contextual awareness and empathy \cite{sauvola2024future}.

The accuracy of proto-personas often created under tight time and resource constraints remains a key concern. It is especially in startups, where access to quality user data and domain specialists is limited \cite{pinheiro2018contribution, wang2025personas_practicioners}. These constraints, combined with concerns about time, cost, and data privacy, often discourage direct user research. Although GenAI has improved productivity in tasks such as coding and documentation, creative RE remains limited \cite{sauvola2024future}. Such tasks require user-centered reasoning and contextual interpretation. Issues like hallucinations, lack of transparency, and trial-and-error in prompt design persist \cite{fan2023large, banh2025copiloting, bazzan2024analysing}. To address these, Bazzan et al. \cite{bazzan2024analysing} emphasize the need for effective prompt engineering to control GenAI behavior.

GenAI can reduce barriers to proto-persona creation by rapidly generating plausible user profiles from contextual prompts. Its natural language abilities help non-specialists more easily express and refine user assumptions. While these benefits make GenAI a promising aid for early-stage design, its empirical application to proto-persona generation remains underexplored \cite{marques2024using, karolita2023use}. LLM-based tools such as PersonaGen \cite{Zhang2023} and AutoPersona \cite{zhang2024auto} indeed facilitate persona creation. However, their use relies on user data, often scarce in early Product Discovery, highlighting the need for practice-based research in real settings. 

To address this gap, we investigate a prompt-engineering approach to support proto-persona generation during Lean Inception (LI). Our goal is to evaluate its effectiveness, user acceptance, and the level of empathy conveyed by the generated proto-personas.

\section{Related Work}

Personas have been utilized across various domains and stages of the design process \cite{salminen2022use}, yet their creation remains a research-intensive task that can strain project timelines and budgets \cite{karolita2023use, karolita2024lessons, wang2025personas_practicioners, losana2021systematic}. Several works advocate for more effective and efficient persona generation methods \cite{karolita2023use, karolita2024lessons}. Tools such as PersonaGen \cite{Zhang2023} and AutoPersona \cite{zhang2024auto} leverage LLMs and knowledge graphs to facilitate persona creation based on user data, though empathy and user acceptance remain under-evaluated in these systems. We explore a prompt-engineering-based approach using LLMs to generate proto-personas during early-stage Product Discovery, where user data is limited, with a specific focus on empathy and user-centered design \cite{bruun2025coordination, karolita2024lessons, enablersBarriersEmpathyGunatilake2024, empathyGunatilake2023}. Our approach complements manual persona generation methods \cite{ferreira2016pathy} by offering a rapid alternative, especially when detailed user data is unavailable, aligning with Liu et al.'s call for structured persona creation processes \cite{liu2022curated}.

There is growing interest in the application of LLMs to SE, including RE \cite{marques2024using, belzner2023large, bazzan2024analysing, banh2025copiloting}. Our work contributes to this emerging area by investigating GenAI applications in the LI phase, addressing gaps such as the empirical evaluation of prompt engineering in RE \cite{arora2023advancing} and the need for valid, scalable, and explainable approaches \cite{belzner2023large, sauvola2024future, kang2025explainitai}. We respond to open challenges raised by Ahmed et al. \cite{ahmed2025} and others, on combining domain-specific knowledge with prompt engineering to produce reliable outputs \cite{fan2023large, nguyen2023generative, jackson2025exploring}. Our method aligns with the "pre-train, prompt, and predict" paradigm \cite{liu2023pre} and extends work like that of Jung et al. \cite{jung2017persona} by shifting the focus from post-deployment persona evolution to early-stage generation. Furthermore, our study evaluates empathy in AI-generated personas—an underexplored area with notable implications for user trust and product alignment \cite{karolita2024lessons, empathyGunatilake2023}. Finally, while prior research applied persona generation to domains like education \cite{zhu2025exploring, gonzalez2024ai} and ethics in computing \cite{gonzalez2024ai}, our contribution centers on practical GenAI integration within Product Discovery.

\section{The approach}

The prompt-engineering-based approach adopted in this case study is a refined version of the method proposed by Leão et al. \cite{leao2024prompt}. It builds upon the prompt patterns cataloged by White et al. \cite{white2023prompt} and OpenAI guidelines \cite{openai2024prompt}. The execution was carried out using the GenAI ChatGPT 4o-mini, in an anonymous tab, accessed via the OpenAI web interface, selected due to its unrestricted usage at the time of the study. Prompts used are available in our GitHub at the Artifacts Availability section, described as "Formal Approach".

The approach (Figure 1) comprises five activities, two of which are derived from prior LI phases: the "Product Vision" and the "Is/Is Not/Does/Does Not" matrix. These inputs are essential for guiding the generation of proto-personas, as they define both the nature and the scope of the product. The first activity contextualizes the LLM about the objective of the approach: creating proto-personas in LI. The second and third activity are responsible for collecting the two artifacts listed above, respectively. The fourth activity establishes the output template of the proto-persona. Finally, the fifth activity incorporates a role in the LLM and asks it to generate the proto-personas.

To reduce hallucinations from the LLM, iterative refinements were made to the original approach. Specifically, the "Context Manager" prompt pattern~\cite{white2023prompt} - allows the user to specify the context for the LLM’s output - was applied during the approach first three activities (Figure 1) to maintain a consistent understanding of the product context. In the fourth activity, a predefined proto-persona template was used to constrain the output according to LI standards\footnote{\label{lean}\url{https://caroli.org/en/lean-inception-4/}}, further mitigating hallucinations. In the fifth activity, the "Persona" prompt pattern - gives the LLM a persona or role to play when generating output - \cite{white2023prompt} was employed to ensure that outputs remained within the UX/UI design domain.

\begin{figure}[h] \centering \includegraphics[width=\linewidth]{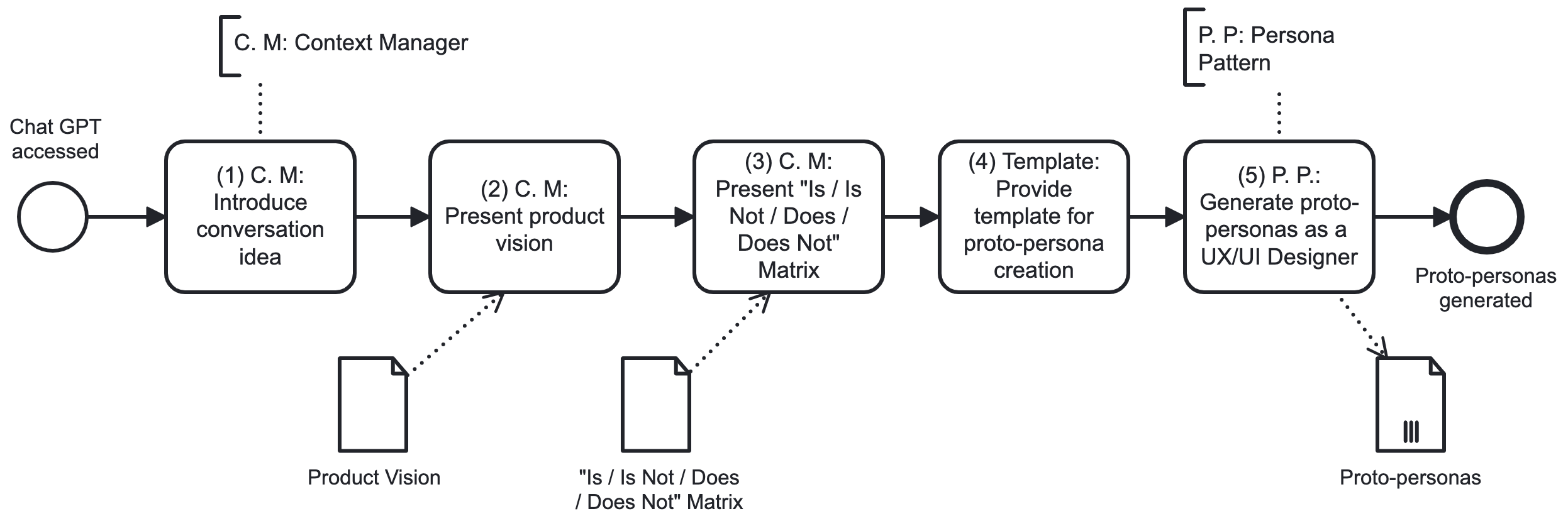} \caption{Refined approach based on Leão et al.~\cite{leao2024prompt}} \label{fig:ourapproach} \end{figure}

\textbf{\underline{Adaptations made in the approach:}} Based on the future work of Leão et al. \cite{leao2024prompt}, two main improvements were made in the fourth and fifth activities. (1) We explicitly defined the proto-persona profile in the “Provide template for proto-persona creation” activity. ChatGPT was previously not instructed to include key attributes as age. Hence, we added age, occupation, marital status, and education, essential demographic data for understanding users \cite{teixeira2022lean}. (2) We modified the “Generate Proto-personas as a UX/UI Designer” activity to always produce at least five proto-personas, following Nielsen’s recommendation to capture most user needs \cite{nielsen1993}.

Additional refinements included improved prompt formatting and instructing the LLM to output only the proto-personas, avoiding introductory or explanatory text. Although not part of the approach itself, visual representations of the final proto-personas were generated, as recommended in the future work of Leão et al. \cite{leao2024prompt}, using ChatGPT 4o-mini after the convergence phase.

\section{Case Study Design}

\subsection{Case and Units of analysis}
\textbf{\textit{The case analyzed in this study}} involves the application of a prompt-engineering-based approach for the construction of proto-personas in the early stages of a Product Discovery within an AI-driven web application project. LI \cite{caroli2017lean} was employed as the primary methodology for discovery, enabling the creation and iterative refinement of proto-personas through a structured four-hour session of divergence and convergence activities. \textbf{\textit{This research is characterized as a single-case study}}, as it investigates a particular phenomenon within a bounded, real-world context. The \textbf{\textit{Unit of Analysis}} is an AI-based web project to support state attorneys and public sector legal professionals by automating tasks such as case retrieval, document analysis, and extraction of legal information. It integrates with multiple external platforms, many of which are proprietary and offer limited interoperability. This system was developed as part of a research and development (R\&D) initiative within a government context, with active involvement from practitioners in the target organization. Beyond technical and domain-specific complexities, the system has the potential for high institutional impact, including public resource savings and improved operational efficiency for legal staff. The legal domain itself presents a high degree of complexity due to regulatory constraints and domain-specific terminology. Combined with the challenges of building and integrating AI-based services, the project demanded a multidisciplinary team: nine software engineers, two computer scientists, four computer engineers, two data scientists, and two legal experts. Product Discovery was conducted through a LI workshop spanning one week, with participants engaging in half-day sessions. This condensed format demanded rapid goal alignment within strict time constraints.

\subsection{Rationale}

Leão et al. \cite{leao2024prompt} highlighted potential challenges in applying prompt-engineering-based approaches to complex projects and advocated for further research. Our unit of analysis presents complexity in terms of business logic and domain knowledge to operate effectively within the project. Also, the work of Leão et al. \cite{leao2024prompt} was conducted months after LI was finished, leading to a necessity for the development of a study that exposes the proposed approach during the actual LI. This case study, therefore, fills critical gaps identified in the literature by (i) evaluating prompt engineering in a high-complexity, domain-specific environment \cite{ahmed2025}; (ii) embedding the intervention within the actual LI process \cite{sauvola2024future}; and, (iii) incorporating empathy as a first-order construct, thereby contributing to ongoing discussions on human-AI collaboration and ethical AI deployment in SE \cite{enablersBarriersEmpathyGunatilake2024}.

\subsection{Research Question and Study's Goal}
Our goal is to \textbf{investigate} the use of a prompt-engineering-based approach \textbf{aiming to} support the efficiency, effectiveness, acceptance, and impact of empathy of this approach when supporting the proto-personas generation task \textbf{from the viewpoint of} researchers and practitioners \textbf{in a context of} a Product Discovery approach, such as LI.  This study supports the following open challenge for Artificial Intelligence (AI) for SE indicated by Ahmed et al. \cite{ahmed2025}: "\textit{How to engineer and combine prompts with domain-specific information to efficiently and effectively address SE tasks}".  To support our goal, we developed the following research questions and metrics:

\textbf{\textit{RQ1: What is the practitioners perception of efficiency of the prompt-engineering-based approach of proto-persona generation?}} \textbf{Rationale: } Efficiency may be defined as the capacity to deliver solutions that meet established requirements, using minimal time, effort, and resources \cite{padua2010measuring, bosch2019efficiency}. It is essential to determine whether GenAI can reduce the time and effort required for proto-persona generation under real-world constraints. The effort reduction is a key factor influencing GenAI adoption in SE tasks \cite{russo2024navigating}. In addition, work pace directly influences the team’s ability to validate hypotheses, discuss functionalities, and proceed confidently toward MVP development \cite{bruun2025coordination}.
\textbf{Metric 1: } Cost in terms of time dispensed to execute the approach, as well as effort to execute it.

\textbf{\textit{RQ2: What is the effectiveness of the prompt-engineering-based approach of proto-persona generation?}} \textbf{Rationale}: The effectiveness must be assessed in terms of how well the results support subsequent activities and align with stakeholder expectations \cite{russo2024navigating}. Practitioners value personas when they clearly inform design and planning decisions, reinforcing that perceived effectiveness is closely tied to downstream impact in the development process \cite{wang2025personas_practicioners}. \textbf{Metric 1:} Effect of the generated proto-personas to the quality of MVP definition during LI.

\textbf{\textit{RQ3: What is the acceptance of the prompt-engineering-based approach of proto-persona generation?}} \textbf{Rationale:} The acceptance of GenAI-based approaches is shaped by users’ perceptions. Evaluating these factors through the Technology Acceptance Model (TAM) \cite{tam1989} constructs is crucial to understand the sociotechnical feasibility of integrating prompt engineering into proto-personas tasks \cite{russo2024navigating}. \textbf{Metric 1: }Perceived Usefulness (PU). \textbf{Metric 2: }Perceived Ease of Use (PEoU). \textbf{Metric 3: }Attitude Toward Use (ATU). \textbf{Metric 4: }Behavioral Intention to Use (BItU).

\textbf{\textit{RQ4: What is the perception of empathy with the proto-personas generated by the prompt-engineering-based approach?}} \textbf{Rationale:} Empathy is the ability to understand and share the feelings of another, and it is a critical human factor in SE  \cite{enablersBarriersEmpathyGunatilake2024}. It influences how well user needs are understood and reflected in design artifacts such as proto-personas \cite{karolita2024lessons}. Given its multidimensional nature—cognitive, affective, and behavioral—assessing empathy perception is essential to determine whether GenAI-generated proto-personas support meaningful user connection and inclusive design \cite{empathyGunatilake2023}. \textbf{Metric 1: } Cognitive empathy. \textbf{Metric 2: }Affective empathy. \textbf{Metric 3: } Behavioural empathy.

\subsection{DATA COLLECTION \& EXECUTION}

\subsubsection{\textbf{Instruments}}
To conduct the study, we created two guiding scripts: one for the participants and another for the researcher. We also developed a form that included a consent agreement, a section for inserting the generated proto-personas, and a set of questions designed to assess the participants' acceptance of the approach and their empathy with the proto-personas, based respectively on the TAM and the dimensions of empathy \cite{empathyGunatilake2023}. All the artifacts listed above are available in the Artifacts Availability (\ref{sec:artifacts_availability}).

We used the 5-Likert-Scale for both the TAM and the empathy questionnaire. All of the instruments above were refined from their versions from the work of Leão et al. \cite{leao2024prompt}. The form's questions were refined to avoid the neutrality present in the work of Leão et al. \cite{leao2024prompt}(e.g, some participants presented neutrality to the question "I believe it is much better to use the approach, rather than a classic proto-persona generation approach." due to the "much better" clause). In addition, we added four more questions to understand the empathy (e.g, "I can understand and interpret the proto-persona, even though I know it was generated by AI."), based on the three dimensions of empathy described by Gunatilake et al.~\cite{empathyGunatilake2023}, of the participants with the generated proto-personas.

Furthermore, we developed an interview protocol based on the guidelines of Runeson \cite{runeson2009guidelines} to guide the fourth step of the data collection. The protocol was based on the TAM and empathy questions. We also added two questions to understand the impact of the approach on the rest of LI to answer RQ2 (e.g, "Did the generated proto-personas help guide discussions and decisions in other stages of LI and align the team on the target user of the product? Why?"). The researchers iteratively reviewed the protocol before the interviews.

\subsubsection{\textbf{Artifacts preparation}}
Before starting the study, we collected the artifacts from the first day of LI (Product Vision and "Is/Is Not/Does/Does Not" LI matrix) and transposed it to plain text format to be prompted in ChatGPT. The transposed artifacts were reviewed by another member from the unit of analysis, so we guaranteed there were no losses or deviations in the content. After that, we inserted the plain-text artifacts in the execution script we provided to each participant.

\subsubsection{\textbf{Participants selection}}

The 19 participants (Table 1) who executed the approach were members of the project's team. Their profiles ranged from highly experienced and specialized professionals to university newcomers in AI and SE. Their academic qualifications cover Computer Science (CS), Computer Engineering (CE), Data Science (DS), Information Systems (IS), Law, and SE. Participants P2 and P19 are affiliated with the state government agency. Among the 17 participants interviewed, nine had prior experience with proto-persona creation, and only six had prior experience with LI. Members with zero months of experience represented 10,53\% of the participants. Table 1 presents the participants' education, area of expertise, and experience (in LI and in projects) to provide context for understanding their perspectives.

\subsubsection{\textbf{Preparation for the study execution}}
We were aware that some participants lacked experience with both LI and project work in general. To mitigate this, we provided prior training to the entire team, explaining what Product Discovery is, its objectives, and its activities. By the time we implemented the approach, all participants had a basic understanding of the Proto-persona creation activity. Still, it is important to emphasize that the contributions of less experienced participants are also valuable, primarily because they bring diverse perspectives on the approach.

\subsubsection{\textbf{Study execution}}
It was divided into \textbf{two parts}: approach execution and interviews. \textbf{The approach execution} started on the second day of the unit of analysis LI, at the proto-persona construction activity. We provided each participant with the execution script, mentioned in the "Artifacts preparation" section, containing detailed instructions for carrying out the approach, along with definitions of key technical terms. During the proto-persona LI \textbf{divergence phase}, the execution was conducted by two researchers. Each participant executed the approach separately, marking the time of execution. The individual approach is crucial to avoid bias and encourage independent input, as recommended by LI book \cite{caroli2017lean}.

\textbf{After the execution}, the participants entered their approach execution time and answered TAM and empathy questions in the previously provided form. The study was then followed by the LI convergence phase \cite{caroli2017lean}, where the participants collaboratively merged their ideas into three final proto-personas. This phase was conducted by one of the researchers as a moderator. The process begins with the moderator asking whether any participant has generated a proto-persona that corresponds to a user profile already familiar to the project. From this discussion, the most relevant proto-personas are elicited, based on each participant’s perspective. The highest-priority proto-persona is then selected, using one participant’s version as a reference. Next, each participant checks whether they produced a similar profile. Guided by the moderator, the group collaboratively refines the reference proto-persona by integrating specific elements from the others. Once finalized, the team proceeds to the next most relevant proto-persona and repeats the process. This cycle continues until five priority proto-personas have been defined.

\textbf{The interviews} started right after LI was over and aimed to provide a qualitative perspective of the approach, describing the reasons behind the TAM answers and the \textbf{notion of the impact of the approach on the LI}. Due to unavailability, two of the 19 participants who executed the approach were not able to attend the interviews. Each interview was conducted by the researchers and resulted in notes and a meeting transcription. Interviews varied between 20 and 47 minutes in duration. We conducted individual semi-structured interviews to reduce bias from senior participants and capture diverse perspectives, given the wide variation in participants’ roles and experience, from novice engineers to domain experts with authority.

\begin{table}[H]
\centering
\caption{Participants’ Profile}
\begin{tabular}{|c|c|c|c|c|}
\hline
\textbf{P} & \textbf{Education} & \textbf{Area} & \textbf{\shortstack{LI \\ Experience (Y/N)}} & \textbf{\shortstack{Project \\ Experience\\(Months)}} \\
\hline
P1  & Master     & DS  & N & 168 \\
P2  & Specialization     & Law & N  & 12  \\
P3  & Doctorate    & AI  & N & 288 \\
P4  & Specialization     & DS  & N & 24  \\
P5  & Student  & SE  & Y & 12  \\
P6  & Graduate     & CE  & Y  & 24  \\
P7  & Graduate     & SE  & Y  & 48  \\
P8  & Graduate     & CE  & N  & 28  \\
P9  & Student  & SE  & N & 24  \\
P10 & Graduate     & IS  & N & 0   \\
P11 & Student  & CS  & N & 12  \\
P12 & Student  & SE  & N & 0   \\
P13 & Graduate     & CE  & Y  & 18  \\
P14 & Student  & SE  & Y & 24  \\
P15 & Master     & CE  & Y & 84  \\
P16 & Student  & SE  & N & 25  \\
P17 & Graduate     & SE  & N  & 16  \\
P18 & Student  & SE  & N & 6   \\
P19 & Specialization     & Law & N  & 3   \\
\hline
\end{tabular}
\end{table}

\subsection{Data analysis}

\underline{Thematic Syntheses.} We conducted an inductive thematic synthesis following a protocol inspired by the guidelines of Cruzes and Dyba \cite{thematicCruzes2011}, employing a codebook-style thematic syntheses approach. The analysis consisted of two parts: (1) manually open coding, and (2) theme extraction, assisted by GenAI. 

Five researchers worked on the analysis, performing the transcription and open-coding individually. We transcribed the recordings using a Python script powered by OpenAI's Whisper model, and subsequently reviewed and corrected the output by cross-referencing it with the original audio. We extracted the preliminary codes and performed iterative conciliation meetings to assure non-repetitive and semantically correct codes.

After establishing a baseline of \textbf{122} codes in the codebook, we used ChatGPT 4o-mini to assist in theme generation, following our qualitative analysis protocol (Section 9) and the guidelines of Roberts et al. \cite{aiThematicRoberts2024} and Yan et al. \cite{aiThematicYan2024}. With these guidelines, we crafted the prompt using “structured task description” and “Input-Process-Output (IPO)” patterns to ensure clarity. Initially, we grouped the codes into 22 unnamed categories; ChatGPT was used to suggest theme names for each, helping reduce researcher bias due to the volume of codes. 

For validation, we reviewed and discussed the themes generated by ChatGPT as a group and chose the most adequate theme based on selection criteria defined in our qualitative protocol. If needed, themes were manually merged or refined by the researchers to better capture nuanced meanings. After extracting the themes, we traced the ones related to each RQ. 

\textbf{\underline{To address RQ1}}, we used the themes and codes from the thematic syntheses related to efficiency. To understand quantitatively the time metric in RQ1, we calculated the average execution time and the sample standard deviation of the proto-persona generation approach and compared it with the time for the same activity in the conventional LI activity (four hour session \cite{caroli2017lean}). 

\textbf{\underline{To answer RQ2}}, we used the themes and codes from the thematic syntheses. Through an Ishikawa diagram, we mapped the effect codes, derived from the "impact on LI questions" in the interviews, and identified possible cause codes that could explain those effects. To validate these relationships, researchers revisited the quotes associated with each code to assess whether they plausibly contributed to the observed impacts on LI. In the Ishikawa Diagram (Figure 2), the codes were organized as causes, grouped under their respective themes as classifications.

\textbf{\underline{To answer RQ3}}, we collected data from the TAM questionnaire and analyzed each statement. It aims to understand if there were trends toward agreement, disagreement, or neutrality. In the case of agreement, we verified if there was partiality or full agreement. Regarding disagreement or neutrality, we discussed hypotheses that could generate this positioning based on the codes extracted in the thematic syntheses. 

\textbf{\underline{For RQ4}}, we used the themes and codes related to empathy with the proto-personas and traced the enablers and barriers to empathy, inspired by the concepts addressed in Gunatilake's work \cite{enablersBarriersEmpathyGunatilake2024} and applied when using the approach.

\section{Results’ Analysis and Discussion }

\subsubsection{RQ1 - The practitioners perception of efficiency of the prompt-engineering-based approach of proto-persona generation}

Through thematic synthesis, we identified the following themes corresponding to the approach’s efficiency: (1) \textbf{Approach Efficiency}; (2) \textbf{Reduction of Effort in Proto-Persona Construction}. We proceed to examine each metric (time and effort) and discuss them in detail. 

From the \textbf{Approach Efficiency} theme, we identified codes evidencing reduced time and increased agility in proto-persona creation. Participants reported time optimization, clarified by P13: \textit{“The issue of one or two days of work was reduced to one hour, so it was very good”}. P3 reinforced this point by highlighting the role of prompt-engineering in expediting the activity: \textit{“Because I think that is the advantage of prompt-engineering, you can accelerate the process greatly”}. Ebert et al. \cite{ebert2023generative} demonstrate time freeing and increased agility in activities performed by LLMs. Furthermore, the results are supported by our collected quantitative data: the average execution time of our approach was 5.94 minutes with a standard deviation of 2.38 minutes.

Participants added that, without the approach, manual proto-persona generation could delay Product Discovery. P16 stated: \textit{“If it were manual, we would have to spend time creating personas, and some stage of the LI would be compromised”}, which was echoed by P15: \textit{“I think we would reach the same result, but it would take much longer and involve much more discussion”}. This perception of reduced time is further corroborated by the TAM questionnaire (to be discussed in RQ3), with tendency of agreement regarding agility of approach. The time saved also permitted the deepening of discussions about strengthening business context and product functionality \cite{bruun2025coordination}. P9 reported: “\textit{And there was enough time for us to discuss and enrich the proto-personas (...) the proto-personas were enough for us to generate the functionality for our MVP}”.

Effort refers to the operational and cognitive work involved in an activity, including communication, discussion, and information integration \cite{padua2010measuring}. Karolita et al. \cite{karolita2024lessons} point out the limitation posed by the high effort required to construct the elements involved in personas. In contrast, Pereira et al. \cite{jackson2025exploring} indicate a reduction in the operational effort of SE activities when applying LLMs to repetitive tasks, highlighting the reduction in cognitive load. By the theme \textbf{Reduction of Effort in Proto-Persona Construction}, we found that the approach lessened participants’ analytical workload, especially in aspects such as creativity, reasoning for diversifying context-appropriate proto-personas, and writing \cite{zhang2024auto}. P15 commented: \textit{"I think that reasoning was much easier in my case, I am a less creative person. (...) was easier than thinking from the beginning”}. P14 added: \textit{“The effort with the prompt process, it is more of a review effort. So, much less tiring, it is much lighter”}. With the approach, the effort previously dedicated to creation was transferred to critical analysis, review, and refinement: activities of higher strategic value in the product discovery process \cite{bosch2019efficiency}. 

Productivity encompasses both the time and effort required to perform an activity \cite{jackson2025exploring}. The \textbf{Approach Efficiency} theme revealed a marked increase in team productivity: P11 exemplified this: \textit{“(...) we did it for 5 [proto-personas]... and if we had to do it manually for 5, sometimes you spend 2 minutes just thinking of a name, it [the approach] makes it much easier”}. Similarly, Belzner et al. \cite{belzner2023large} demonstrated that LLMs can boost efficiency by reducing time and effort, offering an efficient alternative to manual generation. Additionally, participants highlighted the practicality gained from eliminating manual operational tasks. P7 added, \textit{“(...) the approach was already structured, which removed some of the effort of rethinking the prompt”}. By streamlining repetitive tasks, our approach enhanced both productivity and practicality, enabling the generation of more context-specific proto-personas with less manual effort \cite{marques2024using}.

\finding{RQ1}{Our prompt-engineering approach cut proto-persona creation from days to under six minutes, automating repetitive steps and accelerating output. Participants reported increased productivity, richer context, and more strategic discussions.}

\subsubsection{RQ2 - the effectiveness of the prompt-engineering-based approach of proto-persona generation} This section discusses the causes and effects of applying the approach in LI, through thematic synthesis.

One single theme was extracted regarding the effect of the approach in LI: \textbf{Effect of approach on the remainder in Product Discovery}, composed of three codes, each one representing a single effect. Their possible causes were consolidated in an Ishikawa Diagram (Figure 2). Each paragraph below brings the \textbf{effects} and the possible causes that contributed to it.

\begin{figure}[h] 
\centering 
\includegraphics[width=\linewidth]{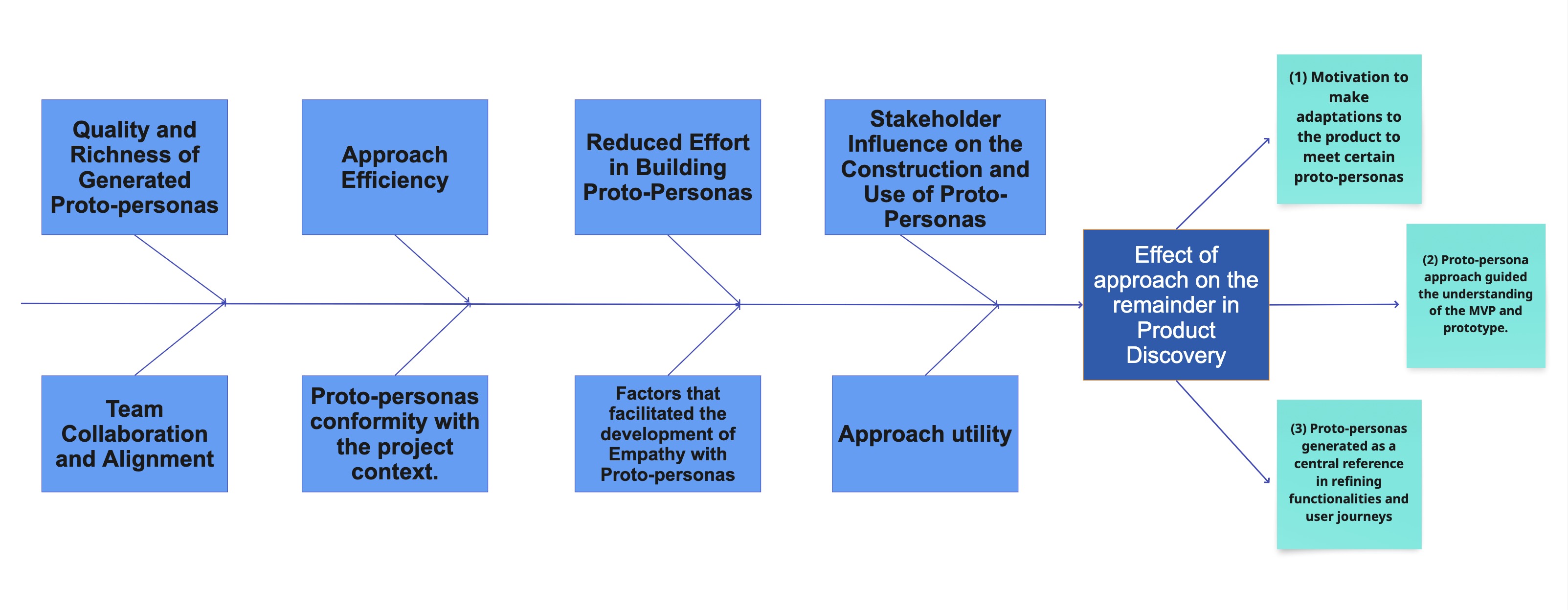} 
\caption{Summarized Ishikawa Diagram}
\end{figure}

\textbf{\underline{Effect 1:} Motivation to make adaptations to the product to meet certain proto-personas}. Developing empathy with the proto-personas (not necessarily using AI) is a factor that contributes to the motivation to help them \cite{karolita2024lessons}. The theme \textbf{Factors that facilitated the development of empathy with proto-personas} reinforces this aspect. P7 expresses it: \textit{“What contributes most to identification is behavior (...) For example, (...) the demanding persona relies on the accuracy of the data. We also feel it, even as researchers}”. This connection helped participants to think about the difficulties of the proto-personas, which culminated in the desire to solve these needs. 

Another aspect that possibly contributed to the motivation to make adaptations was the fidelity of the AI-generated proto-personas to the real-world context—through attributes like education, age, and occupation. This aspect is consolidated by the \textbf{Conformity of the Proto-personas with the Project Context} theme. As P6 notes: “\textit{And it creates that persona with something that you will already deal with on a daily basis, in your routine. And I think that makes you think that it is a real person...}”. This perceived realism fostered empathy and motivation among participants, reinforcing their engagement in designing for these users \cite{McIntosh2024re}.

Finally, we perceived that during the approach there was an encouragement of participation, as reinforced by the theme \textbf{Team Collaboration and Alignment}. P8 comments: “\textit{So people wanted to talk, this desire to share became greater and I think that perhaps in a process without this people would be more withdrawn}”. The fact that the proto-personas were generated by an LLM may have influenced participants’ willingness to critique them. Since the personas were not created by a human team member, participants might have felt more comfortable expressing criticisms \cite{gonzalez2024ai}. In this sense, having greater activity in the discussion of proto-personas is beneficial to better understand the proto-personas \cite{wang2025personas_practicioners}. This better understanding, caused by participation, is a possible cause for the motivation to make adaptations of the product \cite{McIntosh2024re}. 

\textbf{\underline{Effect 2:} Proto-personas generated as a central reference in refining functionalities and user journeys}. The aforementioned alignment with user expectations and context, \textbf{Conformity of the Proto-personas with the Project Context} theme made the AI-generated proto-personas valuable not just for initial understanding but also as concrete references in subsequent LI stages. P1 reflects on how these proto-personas influenced functionality decisions: “\textit{If I am building a system for an older audience... I will be concerned with larger fonts (...) I think the proto-persona helps us there}”. The contextual compliance extended even further, enhancing the dialogue with stakeholders during prototyping. P12 emphasizes: “\textit{I think this is something that the Chat proto-personas got right, because it has something that customers were already talking about}”. The alignment with stakeholder expectations made the proto-personas more actionable, turning them into useful inputs for the definition of screens and functionalities. 

As a factor that orbits the approach, we perceived that team collaboration, \textbf{Team Collaboration and Alignment} theme, creates richer proto-personas, which helps the proto-personas being a more solid reference for later phases of LI. P14 comments: “\textit{(...) when we execute them in a larger group of people (...) we can think of multiple different personas (...) and we can arrive at a satisfactory result}”. Moreover, this better understanding of the final users, consequently, the final solution is better understood \cite{karolita2024lessons}, impacting on the effect 3, as well.

The unit of analysis included stakeholders during LI, though not all persona representatives were consistently present. In their absence, the approach’s proto-personas became central to discussions in later LI stages \cite{alignmentGothelf2012} - \textbf{Stakeholder Influence on the Construction and Use of Proto-Personas} theme. P9 details this aspect: “\textit{Maybe, even because one of the stakeholders was not present (...) so we ended up depending more on the proto-personas}”. Also, P1 talks about this aspect: “\textit{It [the approach] is valid because the stakeholder is not always there with us to imagine these persona}". Thus, while the presence or absence of stakeholders constitutes a peripheral aspect of the approach, it significantly impacted the dependence on proto-personas for defining functionalities and user journeys.

\textbf{\underline{Effect 3:} Proto-persona approach guided the understanding of the MVP and prototype}. Participants reported that the richness of the AI-generated proto-personas, \textbf{Quality and richness of proto-personas} theme, not only broadened their understanding but also introduced new perspectives about the potential users of the application. As P6 explains, “\textit{(...) when using AI, it was able to capture other elements that you might not have thought of at that moment}”. As rich and well-developed personas are crucial for understanding user needs \cite{karolita2024lessons}, these new elements, aligned with the participants’ context, \textbf{Conformity of the Proto-personas with the Project Context} theme, contributed to a more refined understanding of the MVP and the user needs.

Furthermore, Karolita et al. \cite{karolita2024lessons} point out that more efficient methods for initial persona drafts could be valuable in freeing up time. As commented at the RQ1 discussion, the approach efficiency, allowed participants to focus more on understanding and improving the proto-personas, leading to richer insights about the MVP.  Zhang et al. \cite{Zhang2023} also convey that automating initial generation, the effort required from the team is reduced, presumably allowing them to focus on subsequent steps like discussion and refinement. As P9 noted, “\textit{So until we extract the most important information... it would take a lot of time}”. P17 reinforced this by stating, “\textit{I think it would require a lot more conversation [without the approach], discussion, seeing who knows more about the context}”, highlighting that the approach democratized participation—even among those with less initial context. According to P15, this shift in focus enabled deeper thinking: “\textit{Maybe we wouldn’t have even reached the same result, because we would have spent time discussing specific points and not thought about others}”. Ultimately, the reduced effort and increased approach efficiency enabled better proto-personas to emerge, serving as solid foundations for functionality design and a more concrete understanding of the MVP.

Additionally, under the \textbf{Approach Utility} theme, the method’s usefulness as an information base enabled participants outside the domain to better understand and contribute to refining the proto-personas. P14 describes: “\textit{we were able to delegate much more of our time to analyzing the proto-personas, objectively, than actually creating them}”. From this viewpoint, having a prepared foundation for discussion with little effort can facilitate the posterior refinement \cite{Zhang2023}. Participants could better understand the proto-personas, which led to a better consciousness of the final solution \cite{karolita2024lessons}.

Finally, as a complement to the approach, the participants manifested that having stakeholders nearby was a great guide in understanding the solution. This aspect is consolidated in the theme \textbf{Stakeholder Influence on the Construction and Use of Proto-Personas}. The user input could help not to overlook crucial requirements and avoid biases \cite{karolita2023use}. P7 noted: “\textit{What helped a lot was that the attorneys were there at the time. They also identified with the behaviors, with the needs and pointed them out.}” Furthermore, this identification with the proto-personas by stakeholders helps the participants’ perception of their realism and credibility. This is an important factor that should be considered concerning understanding the MVP.

\finding{RQ2}{Generated proto-personas were seen as rich, realistic, and contextually aligned, enhancing team empathy, motivating product adaptations, and supporting key activities like MVP definition and user journey mapping. The approach also fostered collaboration, democratized participation, and was effective when stakeholder access was limited..}

\subsubsection{RQ3 - The acceptance of the prompt engineering-based approach of proto-persona generation} In this section, for each TAM topic, we present the results of the Likert scale and perform an analysis based on the data collected in the thematic syntheses.

\textbf{\underline{Perceived Usefulness (PU)}: Using the approach makes it easier for you to build new proto-personas in your projects}. 100\% of the participants agreed on this statement. Evidence of this tendency towards agreement is the satisfaction with the result of the approach, represented by the themes of \textbf{Quality and Richness of the Generated Proto-personas} and \textbf{Conformity of the proto-personas with the Project Context}. P9 presents his view: “\textit{But, I believe that it generated the standard fields of a proto-persona well enough there (...) I think that gave enough richness for you to be able to understand what that proto-persona}”. These results are valuable since proto-personas should reflect the needs and preferences of persona consumers regarding the information included \cite{karolita2024lessons}.

Another factor was the base for discussion that the approach provides to refine the proto-personas — the usefulness of the approach. Since one of the main functions of proto-personas is to serve as a basis for discussion \cite{macedo2024SupportingUR}, the approach aligns with this intended workflow. P15 supports it: “\textit{And it sets a ground, the basis of what will be worked on, everyone is on the same level from there. (...) So it speeds up discussions, consensus}”.

On the other hand, despite some conformity in the proto-personas, the participants reported some generalization - \textbf{Inconsistencies and Generalizations in Proto-persona Generation} - or partial conformity - \textbf{Partial conformity of proto-personas concerning the project context} - of the LLM regarding the project. P3 brings this frustration: “\textit{...the way it was generated here, they are personas for a system, which analyzes legal documents, but our system is for attorneys, and it is for attorney advisors, so... It has needs that are a little different from those that were raised here}”. P12 discusses the lack of specialization of proto-personas: “\textit{The part about needs, objectives and behavior for everyone (...) it seems like it was the same thing written in a different way.}” Indeed, LLM often struggles with complex tasks involving semantic nuance \cite{russo2024navigating} or lack of nuanced understanding in specialized domains \cite{arora2023advancing} - as the legal - domain. That aspect opposes the novelty presented by the LLM in other studies as Marques et al. \cite{marques2024using}. This LLM limitation may have caused more generic or inconsistent proto-personas.

However, this partial conformity and generalization was softened by the team's collaboration \cite{munch2020ProductDB}, evidenced by the code \textbf{Using the approach in collaboration helps to complement each other's proto-personas}. P14 defends this point: “\textit{...when executed in a larger group of people (...) the execution it worked very well, because we can think of multiple different personas (...) and we can reach a result, a satisfactory result}”.

Furthermore, participants reported some possible improvements to the approach - \textbf{Increments for Approach Improvements} -, generally pointing to the construction of a tool or even greater interactivity in the approach. P9 comments: “\textit{I believe that perhaps a platform, an application like that that already had the basic prompts, basic prompts already inserted, right, in the ChatGPT}”.

\textbf{Using the approach accelerated the proto-persona creation.} 89\% of participants agreed and 11\% were neutral. The tendency for participants to agree is reinforced mainly by the theme of \textbf{Approach Efficiency}. P13 argues: “\textit{The issue of one or two days of work was reduced by one hour, so it was very good}”. This time saving is also related to the theme of \textbf{Approach Usefulness}, more specifically, its use as a basis for discussion. Most participants brought up the argument that having something to start from (the result of the approach) greatly speeds up the refinement of proto-personas. This is coherent, considering that persona creation demands ongoing research to integrate new information \cite{karolita2024lessons}. P12 comments on this fact: “\textit{You have a small idea in your head, but putting it on paper will clarify things and GPT Chat is good for that}”. 

The neutrality present in the responses can be connected to the \textbf{Impression of Approach Redundancy} code when discussions about the proto-persona have already been held with the client before. The participants in neutrality had already been used to a similar project, with some members from the current team, for two years. Therefore, they had the impression that they were reviewing discussions previously held with the client. P15 describes “\textit{Like, maybe, in this specific project, we already had these people mapped out, because the client was always telling us what he wanted to attack. So (...) it seemed like we were repeating the same thing}”. It seems to have been only an impression at the time of execution, as P15 himself says later: “\textit{I think it was more initial, more of the moment}”.

Finally, there was some considerations about non-functional requirement of accessibility, \textbf{Accessibility issues in the approach} theme, where some controversial responses were elicited. On one hand, participants described the lack of accessibility information in the generated proto-personas. 
P16 describes this gap: "\textit{None of the personas had limitations (...) That fact has made it difficult to create empathy or think about accessibility}". This gap of non-functional requirements, as accessibility, is also a problematic mentioned by \cite{zhang2023personagen}. This issue could be addressed by refining the prompts to better guide the LLM in considering non-functional requirements. On the other hand, participants as P7 expressed the contrary: "\textit{We also started to think about accessibility [with the approach]}". This dilema suggests mixed views on coverage of non-functional requirements in LLM created proto-personas. Hence, further investigation on this topic is required, as practitioners recognize the significance of incorporating human-centric factors into personas \cite{wang2025personas_practicioners}.

\textbf{\underline{Perceived Ease of Use (PEoU):} It was easy for me to understand the instructions of the approach. } 95\% of participants agreed and 5\% neither agreed nor disagreed. The agreement perceived at the TAM distribution is reinforced by the themes of \textbf{Ease of Approach Execution} and \textbf{Reduced Effort in Building Proto-Personas}. P7 discourses: "\textit{Since I already had the structured prompt, I just copied it, took the parts from LI and put together the prompt. This made it much easier (...) there was no difficulty}". It confirms the tendency indicated by Leão et al. \cite{leao2024prompt}.

Regarding the 5\% neutrality, by looking at the interview notes, we perceived that the participant had a little confusion while executing the approach. P16 found it hard to reconcile the participant script and the form filling, during the approach execution: "\textit{And I remember that I made a slight confusion (...) So if I could unify everything in one place, it would be easier}". This type of confusion was also a present theme in thematic syntheses, represented by the \textbf{Inexperience in aspects surrounding the approach} and \textbf{Difficulty in building and interpreting proto-personas}. Participants like P19, who came from the legal domain, had limited exposure to the computing field. As P19 noted: "\textit{I think the biggest challenge was that I really didn't know the reality of your world, the more technical side. The names, like you said, I have difficulty knowing.}" Even when using LLMs, such professionals may still face challenges. This lack of experience, combined with the need to manage multiple artifacts during the research, may have contributed to a neutral stance.

However, the thematic syntheses showed another aspects that were not explained by the Likert Distribution, considering Perceived Ease of Use. The participants elicited some prerequisites for using the approach, consolidated in the \textbf{Preconditions for Executing the approach} theme. Aspects as human validation, stakeholder maturity, well defined context and prompts \cite{karolita2024lessons}, known domain by the participants were considered as necessary for the approach execution. Regarding context definition, P9 delves deeper: "\textit{I think the AI is very capable of generating proto-personas (...) as long as you pass on the information it needs for that}". In addition, the theme \textbf{Challenges for Executing the Approach} exposed some concerns of the participants about the approach, as hallucination, data privacy \cite{russo2024navigating}. Also, as pointed out by the work of Leão et al. \cite{leao2024prompt}, the application of the approach in really complex projects. P16 comments: "\textit{But maybe in this case (...) so that it has very unusual personas or personas that are very different from each other (...) Maybe it will hamper reasoning and creativity}". 

\textbf{\underline{Attitude Toward Use (ATU):} Using the approach can be a good approach for generating proto-personas.} 100\% of the participants agreed. The qualitative data reinforces the distribution, mainly through the theme \textbf{Approach Reusability}. Participants, as P7 and P12, advocate that the several options (of proto-personas) that the approach brings are generally helpful for the proto-persona generation. P7 reinforces: "\textit{I don't see why not to use it (...) I always like to have several options}". P12 contributes: "\textit{I don't think there is this scenario that you can't use it, at least to (...) create a base, an initial vision}". The creation of proto-personas is often time consuming \cite{wang2025personas_practicioners} and having a facilitated ground for refinement decreases the initial effort to the participants, as reinforced  by the \textbf{Reduced Effort in Building Proto-Personas} theme.

\textbf{I enjoy using AI tools and processes to support my UX activities.} 89\% of the participants agreed, and 11\% presented neutrality. The agreement is pretty much plausible with the \textbf{Approach Reusability} theme, more specifically with codes such as \textbf{No issues regarding the use of AI}. Participants, in general, were positive about the AI use for the proto-persona creation. P1 describes: "\textit{When using LLM, I didn't see any concerns. We weren't doing anything confidential}". The general agreement colaborate with the increasing adoption of AI-generative tools to create personas \cite{wang2025personas_practicioners}.

Regarding neutrality, participants in this category elicited some concerns in the use of AI, mainly due to data privacy issues and hallucination. P3 describes: "\textit{Well, there are two problems, right? Privacy. But I think that in the proto-persona generation, there's not much of a problem uploading those things to the internet. Unless it's a top-secret project, but in principle, there's not much of a problem. And another problem with LLM is very much related to hallucination}". These two concerns are common barriers to AI adoption \cite{russo2024navigating}, acting as, respectively, Security worries and Quality concerns.

\textbf{\underline{Behavioral Intention to Use (BItU)}: I intend to use the approach whenever possible.} 84\% of the participants agreed and 16\% presented neutrality. Themes as \textbf{Team Collaboration and Alignment} were motivators to the approach's reusability. For instance, P3 comments that people were more aligned with the approach: "\textit{And then, from that point on, everyone was already aligned}". Moreover, the theme \textbf{Factors that increase confidence in the approach} was also a possible cause. P5 describes the benefit of the structured prompt engineering approach: "\textit{... precisely having a very clear approach with a prompt that is done, tested, validated, which gives you security}". The positive intention suggests the compatibility of the approach with the workflow of creating proto-personas \cite{russo2024navigating}. This aspect is also reinforced by the code \textbf{Approach as a Tool to Help Build Proto-personas, Not Automate }.

Regarding neutrality, we observed that the possible cause is a lack of connection of empathy with the generated proto-personas. P12 delves deeper into this aspect: "\textit{But I didn't feel it... (...) I didn't feel that sense of empathy with them}". P14 also describes this frustration: "\textit{but if I need to connect with the Proto-Persona, think from its thoughts (...) I would have more difficulty}". Indeed, proto-persona that transmit empathy are better at conveying authenticity and fostering believability \cite{karolita2024lessons, enablersBarriersEmpathyGunatilake2024}. This gap can hinder the understanding of the end-users \cite{karolita2024lessons}.

\textbf{I would adopt new tools similar to the approach in the future.} 100\% of the participants agreed. By our qualitative data, it is coherent that most participants agree in using new tools similar to our approach. Themes as \textbf{Team Collaboration and Alignment}, \textbf{Ease of Approach Execution} and \textbf{Approach Reusability} reinforce the positivity towards the adoption of similar tools, as P14 comments: "\textit{But I think it's really a method that... I would use again}".

\finding{RQ3}{Results show broad TAM acceptance: most participants found the approach useful, easy to use, and valuable for generating context-rich proto-personas, speeding up design, and supporting collaboration. Despite minor concerns about generalization, prerequisites, privacy, and empathy gaps, participants expressed strong intent to reuse and adopt similar tools.}

\subsubsection{RQ4 - The perception of empathy with the proto-personas generated by the prompt-engineering-based approach} This RQ investigates the aspects of empathy with the proto-personas generated by the approach, based on the three dimensions of empathy: Cognitive, Affective and Behavioral, elicited by Gunatilake et al. \cite{empathyGunatilake2023}. Through the thematic analysis, two central themes focused on empathy were obtained: \textbf{Difficulty in establishing Empathy with the Proto-persona using the Approach} and \textbf{Factors that facilitated the development of Empathy with Proto-personas}. The following discussion links each empathy dimension with the findings from the thematic analysis.

\textbf{\underline{Cognitive empathy}.} The perception of empathy in this scenario is related to how much the participants were able to abstract and interpret the generated proto-personas. It is defined by Gunatilake et al. as "the tendency to understand, or the state of understanding, the internal states of others" \cite{empathyGunatilake2023}. Its presence in the approach is attested by the theme \textbf{Factors that facilitated the development of Empathy with Proto-personas}. In this scenario, given the results presented by the Likert scale, 100\% of the participants demonstrated agreement regarding the facilitation of the understanding and interpretation of the proto-personas generated, through the approach. Given the unilateral positive agreement of the participants, at the core of the cognitive dimension, we have the presentation of two codes facilitating cognitive empathy: (1) \textbf{Template as a whole helped in the interpretation of the proto-persona, not a specific field} and (2) \textbf{Importance of attributes such as age and function for the understanding of the proto-personas}. 

Regarding the first code, participant P17 elaborates: "\textit{... You were able to imagine more tangibly, right, the people you are dealing with there}". Subsequently, the second code contains the following comment from P4: "\textit{... And we were also able to cover the issue of age, the issue of mastery of the tool}". Both address the key points that directly impacted the participants' unanimity regarding the understanding of the generated proto-personas, since these are attributes added after the feedback analysis, provided by the work of Leão et al. \cite{leao2024prompt}. Thus, the increase in these new attributes allowed the generation of more complete proto-personas, which, according to Liu et al. \cite{liu2022curated}, generates greater empathy. 

\textbf{\underline{Affective empathy}.} This dimension of empathy relates to scenarios where participants can see themselves in the proto-persona’s place, defined as “feeling the same affective state as another person” \cite{empathyGunatilake2023}. Affective empathy presented the lowest level of agreement among participants, based on the Likert scale data. Neutrality and disagreement presented values of 32\% and 15\%, respectively, regarding the issue of connection with the proto-persona and seeing themselves as a real person. Together, these values represent almost half of the sample interviewed (47\%), indicating a barrier in developing a connection with the proto-personas.

This barrier is consolidated in the theme \textbf{Difficulty in establishing Empathy with the Proto-persona using the Approach}. Mainly through the codes \textbf{Lack of self-identification with the proto-persona} and \textbf{Lack of affective connection with the proto-persona}, which specifically address the lack of representation during the execution of the approach. One of the reasons for this is the divergence of contexts between the participants and the proto-personas, as commented by participant P3: "\textit{I can't see myself because I don't do those things. It didn't make much sense}".

In this scenario, affective perspective also presents facilitating codes - theme \textbf{Factors that facilitated the development of Empathy with Proto-personas}. An example is the code \textbf{Creation of a connection of feeling when living in a context similar to that of the generated proto-personas}, which transcribes the exact mention of participant P4: "\textit{No, I think that the fact of (...) living with people similar to the proto-personas ends up creating this connection}". This suggests that participants’ prior experience in contexts similar to those of the generated proto-personas fosters affective empathy, aligning with findings by Gunatilake et al. \cite{enablersBarriersEmpathyGunatilake2024}.

\textbf{\underline{Behavioral empathy}.} This dimension has two types: behavior spreading and empathic communication \cite{empathyGunatilake2023}. In the approach execution scenario, the prevailing perception is that of empathic communication, defined as "the intentional behavior that demonstrates cognitive and/or affective empathy towards the other person" \cite{empathyGunatilake2023}. This aspect is also present in both the themes \textbf{Difficulty in establishing Empathy with the Proto-persona using the Approach} and in \textbf{Factors that facilitated the development of Empathy with Proto-personas}. Likert scale responses show participants did not fully feel able to act on the proto-personas, with no one completely agreeing. Moreover, 21\% disagreed that the approach influenced empathic communication.

There is controversy regarding this, since the theme \textbf{Factors that facilitated the development of Empathy with Proto-personas} brings the following facilitating code: \textbf{Identification with the proto-personas based on their behavior}. This theme identifies that the approach allows the participant to get closer to the generated proto-personas through the behavior of the proto-personas described in the LI template. This can be exemplified by the response of participant P13: “\textit{(...) considering the content that the Chat brought, it made me much more related to these personas (...)}”.

However, participants' neutrality was also present in both questions, varying between 10\% and 11\%. The theme \textbf{Difficulty in establishing Empathy with the Proto-persona using the Approach} addresses the lack of motivation to act in favor of the generated proto-personas. This factor becomes explicit in the code \textbf{No motivation to act in any way concerning the proto-persona}, when it brings the following comment from participant P16: “\textit{To have empathy, the persona needs human details (routines, preferences). The AI did not generate this spontaneously”}. This result can be attributed in part to GenAI's limitations in creative and human-centered activities \cite{jackson2025exploring}, particularly concerning behavioral aspects.

\finding{RQ4}{Cognitive empathy was consistently supported, as participants found the proto-personas clear and contextually coherent. In contrast, affective and behavioral empathy yielded mixed results: while some related to behavior-based traits, nearly half struggled to form emotional connections or feel motivated to act, perceiving the personas as too artificial. }

\section{Limitations}

As a qualitative case study, this research presents typical limitations, which we address through the principles of credibility, reflexivity, multivocality, rigor, and transferability \cite{ralph2020empirical}.

\textbf{Limited model scope}. We evaluated a single LLM without comparing alternatives like Gemini-1.5-Pro or LLaMA-2.7, so findings reflect this model’s capabilities and limits. Despite potential model-specific effects on persona quality and perceptions, our goal was to assess the feasibility and usefulness of a lightweight, publicly accessible solution suited to real-world constraints. \textbf{Sample size and data saturation}. We achieved inductive thematic saturation \cite{saunders2018saturation} with 17 interviews, exceeding typical thresholds for identifying dominant themes. Including both experienced and novice practitioners ensured multivocality, enriching perspectives and enhancing analytic depth and trustworthiness.

\textbf{Contextual scope and participant diversity}. This study was conducted within a legal-tech project developed by a government-aligned software team, which may limit transferability to other domains. However, the project's complex socio-technical nature and the diversity of participant backgrounds support theoretical generalization to similarly complex settings. Still, broader validation across contrasting organizational contexts is needed. Additionally, most participants were from the Computing field, with only two from other areas, meaning the evaluation primarily reflects the perspective of technology professionals. \textbf{Researcher reflexivity and potential bias.} As with any qualitative thematic analysis, researcher bias in theme construction and interpretation is a concern. To mitigate this, we followed a rigorous multi-stage protocol inspired by Cruzes and Dyba \cite{thematicCruzes2011}, including structured coding, LLM-assisted synthesis, and collaborative review. All decisions and artifacts were documented for recoverability. While full objectivity is not expected in interpretivist research, we emphasize transparency and critical reflexivity throughout the process.

\section{Implications}

\textbf{\textit{Efficiency and task redistribution.}} The approach notably reduced the time and cognitive load associated with the creation of proto-personas and allowed teams to dedicate more time to strategic discussions and refinement of functionalities. Practitioners can benefit from adopting prompt-engineering to streamline discovery activities, especially in contexts where time constraints are critical.

\textbf{\textit{Empathy limitations in LLM-generated content. }}While cognitive empathy was consistently observed, affective and behavioral empathy were weaker. Participants often struggled to establish emotional connections or motivation to act on the generated personas. These results expose a limitation in the human-likeness of LLM outputs, where capturing emotional nuance remains challenging. 

\textbf{\textit{Importance of contextual fidelity. }}Contextual alignment was critical to the perceived usefulness of generated proto-personas. When personas matched domain vocabulary, roles, and user realities, participants reported greater motivation and clarity during subsequent LI phases. However, LLM-driven generalizations raised concerns. Practitioners should address this by defining input artifacts precisely and embedding contextual constraints in prompts. 

\textbf{\textit{Collaboration and stakeholder mediation. }}The approach effectively supported team collaboration in the absence of direct stakeholder access. Generated proto-personas acted as shared reference points, fostering alignment, participatory refinement, and democratized ideation. These results suggest that prompt engineering can enhance group engagement in early design. 

\section{Conclusion and Future Work}

The approach proved efficient and effective, reducing the time and effort required to create proto-personas while enhancing productivity and practicality. It enabled the generation of realistic, context-aware personas that supported collaboration and informed decision-making, even with limited stakeholder input. Participants positively received the method across all TAM dimensions, with no disagreement reported; neutral feedback was linked to overgeneralization, occasional confusion, or limited emotional connection. Empathy-wise, the approach strongly fostered cognitive empathy, showed mixed results in affective empathy, and revealed opportunities for improvement in behavioral empathy, particularly in strengthening engagement and alignment with user behaviors.

\textbf{Future work.} Explore developing a more intuitive tool to support the approach, integrating image generation, and improving non-functional aspects such as accessibility. Enhancing affective and behavioral empathy remains a challenge, suggesting hybrid strategies like narrative techniques or stakeholder reviews. Adaptive prompting with domain-specific cues may improve contextual fidelity. Further studies should also examine GenAI's role in human–AI co-creation, support collaboration without direct stakeholder access, streamline convergence phases, and test the approach across diverse persona profiles to assess generalizability.

\section{Artifacts Availability}\label{sec:artifacts_availability}

Our artifacts are available in our public GitHub repository\footnote{\url{https://github.com/TungTungTralaleroTralala/generating-proto-personas}\label{tralaleroTralala}}.

\section{Acknowledgments}

The authors thank FUNDECT (Call No. 42/2024) and UFMS for their support.

\bibliographystyle{ACM-Reference-Format}
\bibliography{sample-base}


\end{document}